\documentclass{emulateapj}
\usepackage{apjfonts}
\usepackage{amsbsy}

\newcommand{\up}[1]{{\rm #1}}

\newcommand{\bdv}[1]{\pmb{#1}}

\begin{document}

\title{The Lens Galaxy In PG1115+080 is an Ellipse
\altaffilmark{*}}

\author{Jaiyul Yoo\altaffilmark{1}, Christopher S. Kochanek\altaffilmark{1}, 
Emilio E. Falco\altaffilmark{2}, and Brian A. McLeod\altaffilmark{2}}

\altaffiltext{*}{Based on Observations made with the NASA/ESA 
{\it Hubble Space Telescope}, obtained at the Space Telescope Science 
Institute, which is operated by AURA, Inc., under NASA contract NAS5-26555.}

\affil{$^1$Department of Astronomy, The Ohio State University, 
140 West 18th Avenue, Columbus, OH 43210;
jaiyul@astronomy.ohio-state.edu, ckochanek@astronomy.ohio-state.edu}

\affil{$^2$Harvard-Smithsonian Center for Astrophysics, 60 Garden 
Street, Cambridge, MA 02138; efalco@cfa.harvard.edu, bmcleod@cfa.harvard.edu}

\slugcomment{accepted for publication in The Astrophysical Journal}

\begin{abstract}
We use the structure of the Einstein ring image of the quasar host galaxy
in the four-image quasar lens PG1115+080 to determine the angular
structure of the gravitational potential of the lens galaxy.  We
find that it is well described as an ellipsoid and that the best fit
non-ellipsoidal models are consistent with the ellipsoidal model.  We
find upper limits on the standard parameters for the deviation from
an ellipse of $|a^\up{B}_3/a_0|<0.035$ and $|a^\up{B}_4/a_0|<0.064$.  
We also find that the position
of the center of mass is consistent with the center of light, with an
upper limit of $|\Delta\bdv{x}|<0\farcs005$ on the offset between them.  
Neither
the ellipsoidal nor the non-ellipsoidal models can reproduce the 
observed image flux ratios while simultaneously maintaining a 
reasonable fit to the Einstein ring, so the anomalous flux ratio of
the A$_1$ and A$_2$ quasar images must be due to substructure in the 
gravitational potential such as compact satellite galaxies or stellar
microlenses rather than odd angular structure in the lens galaxy.
\end{abstract}


\section{Introduction}
\label{sec:int}

The scenario of hierarchical structure formation dominated by cold dark matter
has been extremely successful in explaining various observations of large
scale structure (see, e.g., \citealt{sdss,2df,wmap}).  In this scenario, small 
structures form first and merge to build progressively larger structure 
\citep{martin,simon}.  One prediction of these theories, both analytically
and in simulations, is the existence of far more low-mass halos than are 
observed either in the field or as satellites of the Milky Way.  Detailed
studies of small satellite halos show that many survive without being 
tidally disrupted, albeit with some mass loss, and remain as gravitationally
bound {\it substructures} in the halo of the larger galaxy \citep{ben,and}.
Most estimates lead to an abundance of satellites that are inconsistent with
the observed abundances, suggesting that the substructure must consist of
dark satellites in which star formation was suppressed by heating and 
ejecting the baryons \citep{james}.  Unfortunately, dark satellite halos are
difficult to detect.

Fortunately, we have one probe of halo structure that can detect substructure
purely through its gravitational effects -- gravitational lensing  
(see, e.g., \citealt{roger3,book,chris4}). Substructure in the halo of a 
gravitational lens galaxy, whether dark satellites or
simply stars, can modify the observed fluxes of the individual images relative
to those in 
a smooth gravitational potential.  These are most striking for the systems
with anomalous flux ratios between images which should have similar flux ratios
given any smooth potential centered on the primary lens galaxy 
(e.g., \citealt{sch,zhao,chi,neal1,benton}).  For
the case of microlensing by the stars in the lens galaxy, the existence of
substructure is easily verified by the temporal variations of the flux ratio
such as those observed in Q2237+0305 \citep{q2237}. Tests for satellites have
been focused on lensed radio sources that are too large to be affected
by microlensing and little affected by the interstellar medium of the
lens galaxy.  \citet{chris3} have shown statistically at least, that
the flux ratios of lenses show the characteristic pattern of demagnified
saddle point images expected for low optical depths of substructure embedded
in a higher optical depth of smoothly distributed dark matter 
\citep{paul2,chuck5}.
If these anomalies are interpreted as being due
to substructure, the estimated mass fraction in the substructure \citep{neal1}
is consistent with theoretical expectations
\citep{ben,james}. There is some debate 
over the contribution from small halos along the line
of sight relative to those associated with the lens galaxy, with most 
studies favoring a dominant contribution from the lens galaxy (e.g., 
\citealt{rob,jacky}). However,
from the point of view of understanding the halo mass function this is a
moot point because these low mass halos are not observed as either satellites
or in the field with an abundance approaching that predicted by 
the CDM scenario.

There have, however, been several explorations of whether the flux ratio
anomalies can be explained by adding complex angular structures to the
gravitational potential of the primary lens galaxy rather than by adding
substructure \citep{wyn2,chris3,ole}.
In particular, \citet{wyn2} showed that for one particularly
simple model with arbitrary angular structure the fit to the data
can be reduced to a problem in linear
algebra in which the lens constraints, including the flux ratios, can be
modeled to arbitrary accuracy simply by including enough terms.  The model
parameters of this study were problematic as the quasar images needed only 
to be fitted to $0\farcs05$, which is more than ten times the actual 
uncertainties. However, the mathematical observation that 
standard models fitting only the image positions
fail to fit the image fluxes only because the gravitational potentials
are assumed to be roughly ellipsoidal is correct. However, \citet{chris3}
considered models with some additional angular structure and 
compared models constrained by either only the positions of the quasar images
or the positions of the images combined with any additional 
constraints available for the system.  In each case, the large
deviations from an ellipsoidal model that could improve the modeling
of the flux ratios were ruled out by the additional constraints.   

In this paper we explore whether the lens galaxy in the four-image
quasar lens PG1115+080 \citep{wey,courbin,keeton,impey,treu} 
has significant deviations from an
ellipsoidal density distribution.
PG1115+080 is an interesting case because the merging A$_1$/A$_2$ images have
a flux ratio of $\simeq1.6$ rather than the ratio of $\simeq1.0$
predicted by smooth models.
We will use the same models with arbitrary angular
structure considered by Evans \& Witt~(2003, and earlier by 
\citealt{hans,wyn1,chris2,zhao})
but examine the constraints imposed by the Einstein ring image of
the quasar host galaxy in detail.  \citet{chris2}
demonstrated that the shape of the Einstein ring completely encodes 
the angular structure of the potential, so we can quantitatively
measure the deviations of the higher order multipoles of the lens
gravity from the best fit ellipsoid.  In \S\ref{sec:obs} we present the 
{\it Hubble Space Telescope (HST)}
observations on which our models are based, and then briefly review our
models and the fitting procedures in \S\ref{sec:model}. Our main results
are presented in \S\ref{sec:results}, and we summarize them in \S\ref{sec:sum}.

\begin{figure}[t]
\centerline{\epsfxsize=3.5truein\epsffile{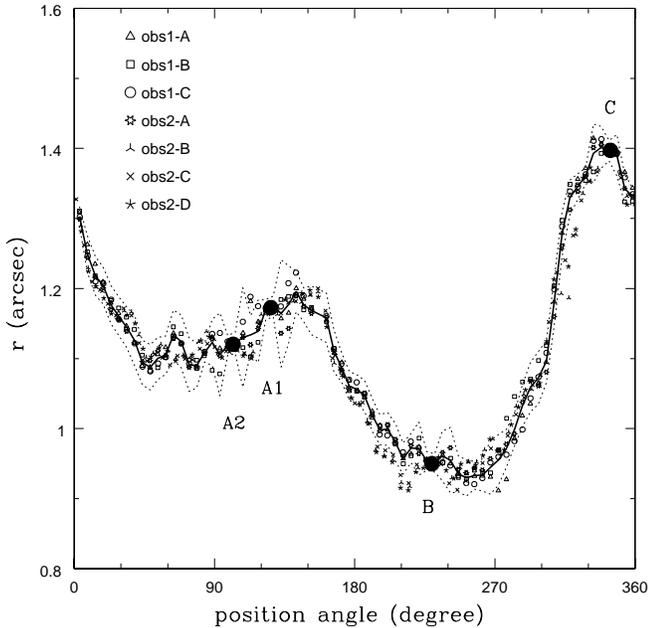}}
\caption{Einstein rings extracted from the data subsamples described in 
\S\ref{sec:obs}. Heavy solid and dotted lines represent the radius of 
the Einstein ring of the final combined
data and its uncertainties, respectively. The four image positions are shown
as filled circles.}
\label{fig:rad}
\end{figure}

\section{HST Observations}
\label{sec:obs}
We use the two $HST$ $H$-band (NICMOS/NIC2/F160W) images of
PG1115+080.  The first observation was a single-orbit (2560s) observation taken
on 1997 November 17 and reported by \citet{impey}.  The second
observation was a four-orbit observation (11700s) 
taken on 2003 June 15. For both
observations, we used four dithered sub-images per orbit to remove hot 
pixels and to improve the flat-fielding.  The data were reduced as 
described in \citet{brian} and \citet{lehar}.
The spacecraft orientation was
$69^\circ$ for the first observation and $-112^\circ$ in the second
observation, so there was little change in the orientation of the
quasar diffraction spikes relative to the lens.

We extracted the ring curves as described in \citet{chris2}. We first fit a 
photometric model to the image including the four lensed quasar images, 
the lens galaxy
and the host galaxy.  This fit provides estimates of the quasar image
fluxes, and these are used to subtract the lensed quasar images from the 
images.
We then select a point close to the center of the lens, and as a function of
position angle $\theta$ we find the radius $r(\theta)$ of the 
peak surface brightness of the Einstein ring and its uncertainties.
The total number of points extracted from an Einstein ring is $N_\up{ring}=71$.
Since the ring is well-detected even in a single image, we extracted
rings from several subsets of the data to explore both statistical
and systematic errors in the data.  We will refer to extractions
from the 1997 data as obs1 and extractions from the 2003 data
as obs2.  For obs1 we extracted the ring using three
different choices of the point about which to extract: obs1-A was
centered on the lens, obs1-B was offset by $0\farcs1$ in the $x$ 
direction of the image and obs1-C was offset by $0\farcs1$ in the $y$
direction.
For the obs2 data we extracted the ring for each of the four
individual orbits ({\it obs2-A, obs2-B, obs2-C and obs2-D}).  For our
final analysis we used a ring extracted from the combined image of the four
individual orbits in obs2.
Figure~\ref{fig:rad} shows $r(\theta)$ for these various
extractions as compared to the final combined model and its
uncertainties. For the image positions and fluxes,
we use the $HST$/NICMOS observations described by \citet{impey}.

\section{Lens Model of PG1115+080}
\label{sec:model}
In this section, we first describe our models for the lens galaxy of 
PG1115+080 and the nearby galaxy group as an external perturber. 
Then we describe our model for the Einstein ring and our approach to fitting
the data.

\subsection{Lens Potential}
\label{sec:lens}
We model the lens galaxy as a scale-free potential with arbitrary angular
structure,
\begin{equation}
\phi(r,\theta)=rF(\theta),~~~~~\kappa(r,\theta)={f(\theta)\over2r},
\end{equation}
where $F(\theta)$ is an arbitrary function of angle $\theta$,
and $f(\theta)=F+F''$ 
\citep{zhao,wyn1,wyn2,chris2,wuc}. These models have a flat rotation curve, 
making them consistent with most estimates of the radial structure of 
lenses (e.g., \citealt{treu,david2}). We expand the angular structure in a
multipole series,
\begin{equation}
F(\theta)=a_0+\sum_{m=2}^\infty\left[a_m\cos m\theta+b_m\sin m\theta\right],
\end{equation}
neglecting the dipole ($m=1$) terms which will be degenerate with a shift
in the source position.
From Poisson's equation, the angular structure of the convergence $\kappa$
is related to that of the potential by
\begin{eqnarray}
f(\theta)&=&a_0+\sum_{m=2}^\infty\left[a_m(1-m^2)\cos m\theta+b_m(1-m^2)\sin
m\theta\right] \nonumber \\
&=&a_0\left[1+\sum_{m=2}^\infty \left(A_m\cos m\theta+B_m\sin m\theta
\right)\right],
\end{eqnarray}
where $A_m=(1-m^2)a_m/a_0$ and $B_m=(1-m^2)b_m/a_0$ for $m\geq2$.
We are interested in the deviations of the potential from an ellipsoidal
surface mass density,
\begin{equation}
f(\theta)\propto\left[1-(1-q^2)\cos^2\theta\right]^{-1/2}
\propto1+\sum_{m=1}^\infty a_{2m}^q\cos2m\theta,
\label{eq:ellipse}
\end{equation}
where $q$ is the axis ratio of the ellipse with coefficients
$a_{2m}^q$ to distinguish them from the coefficients $a_{2m}$ of the
general potential. We will look at deviations of the gravitational potential
from an ellipsoid defined by the $a_0$, $a_2$ and $b_2$ coefficients. In this
model, the quadrupole of the ellipsoid is defined by
\begin{equation}
a_2^q=-{3\over a_0}\left(a_2\cos2\theta_L+b_2\sin2\theta_L\right),
\end{equation}
for axis ratio $q$ and major axis orientation
$\theta_L=0.5\tan^{-1}(b_2/a_2)$.
The remainder of the coefficients $a_{2m}^q$ for $m\geq2$ of the ellipsoidal 
model are defined by the standard multipole expansion of 
equation~(\ref{eq:ellipse}).
We describe the rest of the model by the deviations $\Delta a_m$
and $\Delta b_m$ ($m\geq3$)
from the ellipsoid. Thus, our overall model for the primary lens is
\begin{eqnarray}
F(\theta)&&=a_0+a_2\cos2\theta+b_2\sin2\theta \nonumber \\
&&+\sum_{m=2}^\infty{a_0\over1-4m^2}a_{2m}^q\cos2m(\theta-\theta_L) \nonumber\\
&&+\sum_{m=3}^\infty\left[\Delta a_m\cos m\theta+\Delta b_m\sin m\theta\right],
\end{eqnarray}
where $a_0$, $a_2$, $b_2$, $\Delta a_m$ and $\Delta b_m$ are model parameters.
The deviations in the mass density from the ellipsoid are
$\Delta A_m=(1-m^2)\Delta a_m/a_0=a^\up{B}_m/a_0$ and 
$\Delta B_m=(1-m^2)\Delta b_m/a_0=b^\up{B}_m/a_0$ for $m\geq3$ where
$a^\up{B}_m$ and $b^\up{B}_m$ are the standard coefficients that describe
the deviations of the isophotes of elliptical galaxies from
ellipsoids (e.g., \citealt{iso1,iso2}). Since the lens galaxy is 
fairly round
$q_L\simeq1$, we expand the ellipsoid to order $m=5$ and usually consider 
deviations for $m=3$ to $m=5$ beyond which the coefficients $a^q_{2m}$ are
negligible and the perturbations are well below the noise.

\subsection{The External Perturber}
When modeling gravitational lens systems, it is important to include
an independent external shear that can be misaligned with the light 
distribution \citep{chuck1}. 
In the case of PG1115+080, the dominant tidal perturbations are associated
with the group to which the lens galaxy belongs (e.g., \citealt{impey}).
The group is composed of two luminous spirals and several smaller galaxies
with a luminosity-weighted centroid $\sim20\arcsec$ from the lens.
Since the higher order perturbations of the group beyond an external 
shear are quantitatively important, we need to generalize the models of
\citet{chris2} to include these higher order terms.

We model the group of galaxies as a singular isothermal sphere (SIS), 
$\phi_\up{ext}=b_g|\bdv{r}-\bdv{r}_g|$, where $b_g$ is 
the Einstein radius of the
group and $\bdv{r}_g$ is the position vector of the group relative to the
lens center. If the group is far away from the lens in the sense that
$b/r_g\ll1$, then we need to include only the quadratic terms in an expansion
of the potential,
\begin{equation}
\phi_\up{ext}(r,\theta)\simeq-{1\over4}{b_g\over r_g}r^2-{1\over4}
{b_g\over r_g}r^2\cos2(\theta-\theta_g)+
\mathcal{O}\left({b_g\over r_g^2}r^3\right).
\label{eq:ext}
\end{equation}
The first term provides the {\it mass sheet degeneracy} and affects only
the time delays \citep{emilio,dege}. The second term
is the tidal shear, $\gamma=b_g/2r_g\simeq0.1$ that is the dominant external
perturbation \citep{paul}. However, in PG1115+080 the higher order terms are 
significant, so we must use a more complex model. We approximate the 
potential using only linear and quadratic terms in radial dependence,
\begin{eqnarray}
\phi_\up{ext}(r,\theta)&&\simeq r\sum_{m=1}^\infty a_m^g\cos m(\theta-\theta_g)
+r^2\sum_{m=1}^\infty b_m^g\cos m(\theta-\theta_g) \nonumber \\
&&\equiv rF_\up{ext}(\theta)+r^2G_\up{ext}(\theta),
\end{eqnarray}
because it has the advantage that the structure of the Einstein ring can be 
analytically calculated even when the potential has arbitrary angular 
structure. The quadratic term of the potential $G_\up{ext}(\theta)$
is added to yield a better approximation to 
the radial deflections of a true SIS model. We determined
the coefficients
$a_m^g$ and $b_m^g$ for the external perturber by minimizing the
difference in the deflection of our approximation from that of a true SIS 
near the Einstein ring of PG1115+080. 
We need to expand $\phi_\up{ext}$ only for $m\leq3$, and
the rms deflection difference between the resulting model and a true SIS 
model at
$r_g\simeq10\arcsec$ from the lens is only $0\farcs0005$ over a 
$1\arcsec$ annulus encompassing the ring.
This is significantly less than our smallest astrometric errors ($0\farcs005$).

\subsection{Einstein Ring Models}
\label{sec:ring}
We model the Einstein ring using an extension of the theory developed by
\citet{chris2} for our more general potential,
$\phi(r,\theta)=r\left[F(\theta)+F_\up{ext}(\theta)\right]+r^2G_\up{ext}
(\theta)$.
When the source is extended, the tangentially stretched images of the source
form an Einstein ring. If we assume that the source has a
monotonically decreasing surface brightness, 
the Einstein ring radius relative to the lens is simply 
\begin{equation}
r(\theta)={\bdv{h}\cdot\bdv{S}\cdot\bdv{t}+\bdv{u}_0\cdot\bdv{S}\cdot
\bdv{t}\over\bdv{t}\cdot\bdv{S}\cdot\bdv{t}},
\end{equation}
where $\bdv{h}\equiv(F+F_\up{ext})\hat{\bdv{e}}_r+(F'+F'_\up{ext})
\hat{\bdv{e}}_\theta$, the source plane tangent vector is $\bdv{t}=\bdv{M}^{-1}
\cdot\hat{\bdv{e}}_r=(1-2G_\up{ext})\hat{\bdv{e}}_r-G'_\up{ext}
\hat{\bdv{e}}_\theta$, $\bdv{M}^{-1}$ is the inverse magnification tensor,
the source center is $\bdv{u}_0$, 
and $\bdv{S}$ is the two-dimensional shape tensor of the source
(see \citealt{chris2} for the details).
The only relevant parameters of the source are its axis ratio $q_s$
and the position angle of its major axis $\theta_s$.

\subsection{Model Fitting}
We use a simple $\chi^2$ statistic for the goodness of fit of
a model described by parameters $\bdv{p}$. The astrometric constraints consist
of the observed Einstein ring radius $r(\theta_i)$ and the four source 
positions $\bdv{u}_i$. We divide the estimate into the
goodness of fit to the astrometric constraints $\chi^2_\up{ast}$ and
the flux constraints $\chi^2_\up{flux}$. The fit statistic for the astrometric
constraints is
\begin{equation}
\chi^2_\up{ast}(\bdv{p})=
\sum_i^{N_\up{img}}{|\left[\bdv{u}_0-\bdv{u}_i(\bdv{p})\right]M_i|^2
\over\sigma_{i,\up{img}}^2}+\sum_i^{N_\up{ring}}{|r(\theta_i)-r(\theta_i;
\bdv{p})|^2\over\sigma_{i,\up{ring}}^2},
\label{eq:chisq}
\end{equation}
where $M_i$ is the magnification of image $i$, and the $\sigma_i$'s 
are the uncertainties in either the image positions or the Einstein
ring radii. The fit statistic for images is approximately correct because
of the magnification weightings -- the need to analyze models quickly in a 
large parameter space precluded
using the exact image plane fit statistic.
We can also constrain the model using the image fluxes, although
fluxes can be contaminated by systematic errors, dust extinction,
microlensing by stars, or substructure. For parity-signed fluxes $f_i$,
we add
\begin{equation}
\chi^2_\up{flux}(\{\bdv{p},f_s\})=\sum_i^{N_\up{img}}{|f_i-M_i(\bdv{p})f_s|^2
\over\sigma_{i,\up{flux}}^2},
\end{equation}
where the source flux $f_s$ is an additional free parameter of the fit.

\begin{figure}[t]
\centerline{\epsfxsize=3.5truein\epsffile{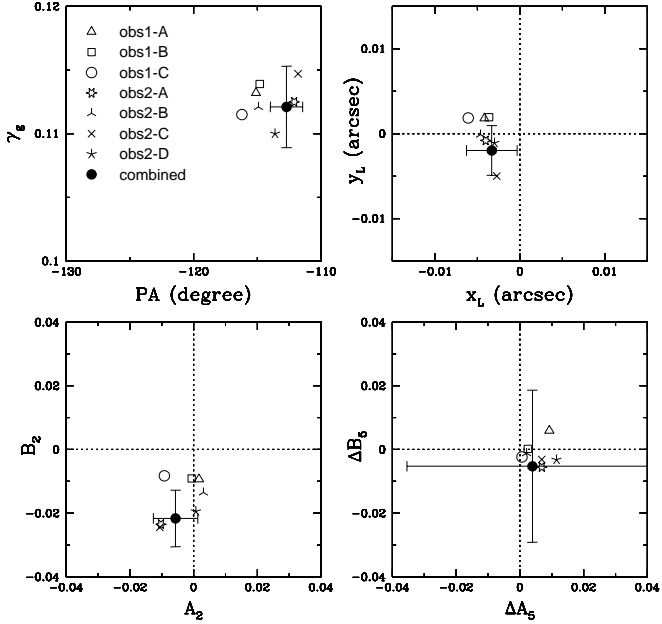}}
\caption{A check for systematic errors by comparing the results for the
rings extracted from the data
subsamples listed in the top left panel and discussed in
\S\ref{sec:obs}. For illustration,
we show the shear and PA of the group ({\it top left}),
the lens galaxy position ({\it top right}), the quadrupole moment of the 
density distribution ({\it lower left}) and the $5\theta$ moments
({\it lower right}). The formal uncertainties in parameters are only shown
for the combined data. For statistical uncertainties, the scatter in the
subsamples should be twice the errorbar for the combined data.}
\label{fig:system}
\end{figure}

We start by fitting the astrometric constraints with a fiducial 
ellipsoidal model that has 12 free 
parameters; the source position is $\bdv{u}_0=(u_x$, $u_y)$, the axis ratio 
$q_s$ and major-axis position angle $\theta_s$ of the source, the shear 
amplitude and the position of the external perturber $\{\gamma_g$, $r_g$, 
$\theta_g\}$, the ellipsoidal
lens potential $\{a_0$, $a_2$, $b_2\}$, and the lens center 
is $(x_L,~y_L)$. Once we have estimated the goodness of fit for this model,
we can add the parameters for the deviations from ellipsoidal symmetry
or $\chi^2_\up{flux}$, and examine the change in the goodness of fit and the 
parameters.

We minimize the $\chi^2$ of a given model and estimate the parameter 
uncertainties using Levenberg-Marquardt method (e.g., \citealt{nr}).
We ensure that we have found a genuine $\chi^2$ minimum by repeating the 
minimization with a range of initial parameters and by checking the existence
of the minimum with the downhill simplex
method. We tested the code on a range of synthetic lenses with Einstein rings
including both ellipsoidal and non-ellipsoidal models.

As a last step before presenting our results, we fit rings extracted from 
subsamples of the data as a check for systematic errors, and the level of our
random errors. We consider the standard ring from obs1
extracted in three different positions relative to the lens galaxy 
({\it obs1-ABC}) and the rings extracted from the four individual orbits of 
obs2 ({\it obs2-ABCD}). In each case the lens was re-modeled as part of the
extraction procedure. Unfortunately, the rotation of the
PSF between the two observations is negligible (see \S\ref{sec:obs}),
so comparisons of the data sets provide little constraint on systematic errors
arising from the PSF.
Figure~\ref{fig:system} compares the results for these seven data sets for a
range of variables using the full non-ellipsoidal models. 
The main point to note is that the parameter estimates are mutually 
consistent. Also note that the distributions of the parameter estimates are 
consistent with the formal statistical uncertainties, except for the 5th order 
deviations where they are smaller.
For the remainder of the paper we consider only the combined data.

\begin{deluxetable*}{lccccc}
\tablewidth{0pt}
\tablecaption{Best-Fit Models}
\tablehead{\colhead{} & \multicolumn{2}{c}{Ellipsoidal Model} & \colhead{} &
\multicolumn{2}{c}{Non-Ellipsoidal Model} \\ 
\cline{2-3} \cline{5-6} \\
\colhead{Parameter} & \colhead{Astrometry Only} & \colhead{Astrometry+Flux} 
& \colhead{} & \colhead{Astrometry Only} & \colhead{Astrometry+Flux} }
\startdata
$q_s$ & 0.69(02) & 0.69(02) & & 0.70(02) & 0.68(03) \\
$\theta_s$(degrees) & $-$32.0(2.7) & $-$30.9(2.6) & & $-$33.4(4.6) & 
                      $-$30.9(3.6)\\
$\gamma_g$ & 0.116(01) & 0.112(01) & & 0.112(03) & 0.112(02)\\
$r_g$(arcsec) & 10.8(1.4) & 12.4(1.8) & & 16.9(12.5) & 12.5(7.1)\\
$\theta_g$(degrees) & $-$113(00) & $-$116(00) & & $-$113(01) & $-$116(01)\\
$x_L$(arcsec) & 0.001(02) & 0.005(02) & & $-$0.003(03) & 0.008(03)\\
$y_L$(arcsec) & 0.001(01) & 0.001(01) & & $-$0.002(03) & 0.004(03)\\
$q_L$ & 0.97(02) & 0.98(01) & & 0.96(03) & 0.99(02) \\
A$_1$/A$_2$ & 0.97(03) & 0.94(02) & & 1.00(13) & 0.93(06) \\
$\theta_L$(degrees) & $-66.1$(7.8) & 68.4(4.9) && $-$52.4(9.2) & 61.8(26.1) \\
$a_0$ & 1.15(00) & 1.15(00) & & 1.15(01) & 1.15(00)\\
$10^3a_2$ & $4.1(2.0)$ & $-3.2(0.9)$ & & $2.2(2.7)$ & $-1.0(1.7)$\\
$10^3b_2$ & $4.5(1.1)$ & $3.0(0.6)$ & & $8.3(3.4)$ & $1.5(1.5)$\\
$10^3\Delta a_3$ & $\equiv0$ & $\equiv0$ & & $-1.4(4.6)$ & $0.4(1.7)$\\
$10^3\Delta b_3$ & $\equiv0$ & $\equiv0$ & & $0.8(7.2)$ & $0.5(2.7)$\\
$10^3\Delta a_4$ & $\equiv0$ & $\equiv0$ & & $-0.1(4.6)$ & $-0.7(1.3)$\\
$10^3\Delta b_4$ & $\equiv0$ & $\equiv0$ & & $0.3(5.7)$ & $-0.3(1.2)$\\
$10^3\Delta a_5$ & $\equiv0$ & $\equiv0$ & & $-0.2(1.9)$ & $0.4(1.6)$\\
$10^3\Delta b_5$ & $\equiv0$ & $\equiv0$ & & $0.3(1.1)$ & $0.1(0.7)$\\ 
\hline 
$\chi^2_\up{ring}$ & 35.6 & 38.1 & & 29.3 & 41.6\\
$\chi^2_\up{image}$ & ~0.2 & ~5.4 & & ~0.1 & ~5.4\\
$\chi^2_\up{lens}$ & ~0.1 & ~2.8 & & ~1.6 & ~9.2\\
$\chi^2_\up{flux}$ & $-$ & 45.4 & & $-$ & ~28.7 \\
$\chi^2_\up{tot}$ & 36.0 & 91.7 & & 31.1 & ~84.8 \\
\enddata
\tablecomments{The best fit ellipsoidal and non-ellipsoidal
($m=5$) models of PG1115+080, fitting either only the astrometric 
constraints or the astrometric constraints and the observed flux ratios.
All the angles are standard position angles while the lens 
position and the coefficients of the lens potential are calculated in 
Cartesian coordinates in which $x$-direction is toward west.
The errors are shown in parentheses.}
\label{tab:best}
\end{deluxetable*}

\begin{figure}[b]
\centerline{\epsfxsize=3.5truein\epsffile{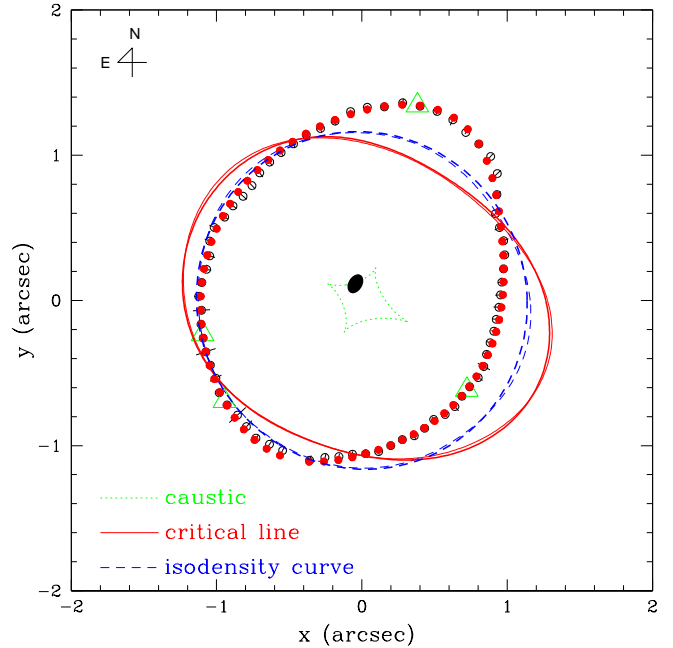}}
\caption{The best fit ellipsoidal ({\it heavy lines}) and $m=5$ non-ellipsoidal
({\it light lines}) models for the ring based on the astrometric 
constraints. The observed positions of the Einstein ring and the lensed images
are represented as open circles and triangles, respectively. The line sizes of
the open circles are the positional uncertainties of the Einstein ring radius 
while the uncertainties in the image positions ($0\farcs003$) are omitted. 
The filled circles represent 
the Einstein ring positions of the best fit ellipsoidal
model, while the filled ellipse 
at the center shows the predicted position, ellipticity and position angle 
of the source. The heavy solid and dashed lines show the critical line
({\it solid}) and the isodensity curve ({\it dashed})
of the ellipsoidal model, while the corresponding curves of the best-fit 
non-ellipsoidal model are shown with the light lines.}
\label{fig:model}
\end{figure}

\section{Results}
\label{sec:results}
In this section we address three issues. First, is the lens consistent with
an ellipsoid when we fit 
only the astrometric constraints? Second, is the center
of mass consistent with the center of light? Third, are any of these
conclusions changed if we try to fit the flux ratios as well?

\subsection{Is the Lens an Ellipsoid?}
We start by fitting only the astrometric constraints, first with a purely
ellipsoidal model, and then adding deviations. The results are presented
in Table~\ref{tab:best}. An ellipsoidal model with a very round lens 
($q_L\simeq1$), a group located at a position angle of 
$\theta_g\simeq-113^\circ$ and at a distance of $r_g\simeq11\arcsec$ provides
a reasonable fit to our data and is consistent with the earlier models
using a singular isothermal ellipsoid for the lens
\citep{paul,keeton,courbin,impey,zhao,chris3}.
The total $\chi^2_\up{tot}\simeq36$ 
for 69 degrees of freedom means that we have somewhat overfit 
the data and/or overestimated the uncertainties by 38\%.
Figure~\ref{fig:model} shows the best fit ellipsoidal model using
only the astrometric constraints. 

Next, we fit the data using a sequence of non-ellipsoidal models to 
investigate how much the lens can deviate from ellipsoidal. We did this by
sequentially including higher order poles starting with $m=3$ ($\Delta a_3$,
$\Delta b_3$), then $m=4$ and finally $m=5$. For comparison with the 
ellipsoidal model, we show the critical line and the isodensity curve of the 
best fit $m=5$ non-ellipsoidal model in Figure~\ref{fig:model}, and
we present their final parameters in Table~\ref{tab:best}.
It is particularly interesting to note that all the higher
order deviations are consistent with zero.

We summarize the resulting $\chi^2$ of each model in Table~\ref{tab:dof}.
We use the F-test to estimate whether the new parameters significantly 
improve the fit, and we find that none of the new variables significantly 
improves the models. In many cases, adding the new variables even raises the
$\chi^2$ per degree of freedom.
All the best fit non-ellipsoidal models are consistent with the best fit 
ellipsoidal model. In short,
the constraints on the angular structure from the Einstein
ring force the lens of PG1115+080 
to be consistent with an ellipsoidal density.
The isophotal deviations of the best fit non-ellipsoidal models are
$\sim10^{-3}$ for both $a^\up{B}_m$ and $b^\up{B}_m$ ($m\leq5$) somewhat
smaller than the values of 
$a^\up{B}_3/a_0=0.009$, $a^\up{B}_4/a_0=-0.004$ for $H$-band, and
$a^\up{B}_3/a_0=-0.015$, $a^\up{B}_4/a_0=-0.004$ for $I$-band 
found by \citet{wyn2}.

In our first models, we constrained the center of mass to agree with the
observed lens position and its formal uncertainties of $0\farcs003$.
\citet{wyn2} were concerned that the center of light and mass may differ,
and allowed the lens center enormous freedom to move relative to the observed
position (0\farcs05). For our next experiment, we drop the constraint on the 
lens position and re-optimize the models. We find a lens position of 
($-0\farcs001\pm0\farcs002$,~$-0\farcs001\pm0\farcs002$)
compared to the measured position of 
($0\farcs000\pm0\farcs003$,~$0\farcs000\pm0\farcs003$) or
a net difference of $0.3\sigma$ for the RA and $0.4\sigma$ for the Dec.
The center of mass of the lens is essentially identical to the center of light.
The observed lens galaxy of PG1115+080 has an axis ratio $q\simeq0.9$
and a position angle $\theta\simeq-70^\circ$ \citep{impey,castle}. The 
best-fit axis ratio $q_L$ of the mass is rounder than the light while
the position angle $\theta_L$ is aligned with the light for both the
elliptical and non-elliptical models.

\begin{deluxetable}{cccccccccc}
\tablewidth{0pt}
\tablecaption{F-Tests of The Additional Degrees of Freedom}
\tablehead{
\colhead{} & \multicolumn{4}{c}{Astrometry only (A)} & \colhead{} &
\multicolumn{4}{c}{Astrometry+Flux (F)} \\ 
\cline{2-5} \cline{7-10} \\
\colhead{Model ($m$)} & \colhead{$N_{par}$} & \colhead{dof} & 
\colhead{$\chi^2_{dof}$} & \colhead{P(\%)} & \colhead{} & \colhead{$N_{par}$} 
& \colhead{dof} & \colhead{$\chi^2_{dof}$} & \colhead{P(\%)}}
\startdata
0 & 12 & 69 & 0.522 & 100  & & 13 & 72 & 1.273 & 100  \\
3 & 14 & 67 & 0.474 & 69.1 & & 15 & 70 & 1.310 & 90.5 \\
4 & 16 & 65 & 0.488 & 78.7 & & 17 & 68 & 1.301 & 92.7 \\
5 & 18 & 63 & 0.493 & 82.0 & & 19 & 66 & 1.285 & 96.6 \\ 
\enddata
\tablecomments{Using astrometric constraints only, models with higher order 
deviations (A$m$) for order $\cos m\theta$
are compared to the fiducial ellipsoidal model (A0). 
When flux constraints are added, models (F$m$) for order $\cos m\theta$
are compared
to F0. The dof column gives the number of degrees of freedom and the
$\chi^2_{dof}$ column gives the $\chi^2$ per degree of freedom.
The probability P gives the F-test probability that 
a given non-ellipsoidal model is consistent with ellipsoidal model. 
A non-ellipsoidal model with lower P implies a more significant
improvement relative to the ellipsoidal model --
none of the improvements is significant.}
\label{tab:dof}
\end{deluxetable}

\subsection{The Flux Ratio Anomalies}
Because we believe the anomalous flux ratios are due to substructure 
(satellites or stars) rather than a problem in the lens model, we have so 
far neglected the flux ratios as a model constraint.
We now investigate the changes in the
lens potential when the flux constraints are added. The best
fit parameters of an ellipsoidal model with flux constraints are 
also shown in 
Table~\ref{tab:best}. There is little change from the parameters of the
ellipsoidal model based only on the astrometric constraints. 
As expected, most of the contribution to the fit statistic $\chi^2_\up{flux}$
comes from the merging A$_1$/A$_2$ image pair showing the well-known flux 
anomaly. To reproduce the observed
flux ratios, the critical line should be either distorted or moved closer to
the brightest image so that the flux ratio becomes higher than 
predicted by ellipsoidal models. In the \citet{wyn2}
models, this was possible
because the quasar and lens positions could be shifted by far more than their
actual uncertainties. However, with the correct QSO uncertainties and the 
Einstein ring constraints, especially on the angular
structure of the lens potential, the best fit model with the flux constraints
largely coincides with that using the astrometric constraints only. 

We then added the higher
order deviations from an ellipsoid
and fit the data with non-ellipsoidal models. Since the
magnification, $M\propto F''\propto m^2a_m$, is more sensitive to the higher 
order multipoles, the best-fit non-ellipsoidal models produce 
a better fit to the flux ratios, but it is still impossible to 
significantly improve flux ratios while simultaneously maintaining a good fit
to the Einstein ring. The best fit model with flux constraints is still 
consistent with the ellipsoidal model. The fit statistics and the
F-test results are shown again in Table~\ref{tab:dof}.
Note that the major axis of the mass is misaligned with the light when
we use the flux ratios as a model constraint.

\section{Summary}
\label{sec:sum}
We modeled the lensed quasar images and the Einstein ring formed from its
host galaxy in the lens PG1115+080 using a scale-free potential with arbitrary
angular structure. The best fit ellipsoidal model is consistent with 
previous results, and provides a reasonable fit to the astrometric constraints
of both the Einstein ring and the images. Non-ellipsoidal models are 
constructed by adding higher-order deviations from the ellipsoidal model,
but none of the additional degrees of freedom improves the models. 
The best-fit non-ellipsoidal models are still consistent with the 
best fit ellipsoidal model. We also find that the center of mass of the lens
is consistent with the measured center of light even when this is not imposed
as a constraint.

When we try to fit the fluxes as well, 
including the anomalous A$_1$/A$_2$ flux ratio,
the best-fit non-ellipsoidal models still fail to
match the observed flux ratios and are still consistent with the ellipsoidal
model. We conclude that the suggestions that complex angular structure in the
lens galaxy can explain the anomalous flux ratio in PG1115+080 are incorrect
-- the galaxy is indistinguishable from an ellipsoid. This result
does not address
the problem of whether the anomaly is due to microlensing or satellites
and whether the source of substructure is in the lens or along the line of
sight. The predicted A$_1$/A$_2$ flux ratio for the astrometry only models
is $0.97\pm0.03$ (see Tab.\ref{tab:best}). There is some evidence that the 
emission line flux ratios are closer to this value \citep{pop}, suggesting
that the anomaly is due to microlensing rather than satellite despite the lack
of time variations in the ratio.
Since Einstein rings are relatively common in $H$-band $HST$ 
observations of lenses, we should be able to test the ellipsoidal hypothesis
in many additional systems.

\acknowledgments
This research has been supported by the NASA ATP grant NAG5-9265, and
by grants HST-GO-7495 and 9375 from the Space Telescope
Science Institute, which is operated by the Association of
Universities for Research in Astronomy, Inc., under NASA
contract NAS 5-26555.

\end{document}